\documentclass[12pt]{article}
\usepackage{a4wide}
\usepackage{epsfig}
\usepackage{amsmath}

\newcommand{\half}{{\textstyle\frac{1}{2}}}

\newlength{\absize}
\catcode`@=11
\def\citer{\@ifnextchar
[{\@tempswatrue\@citexr}{\@tempswafalse\@citexr[]}}

\def\@citexr[#1]#2{\if@filesw\immediate
  \write\@auxout{\string\citation{#2}}\fi
  \def\@citea{}\@cite{\@for\@citeb:=#2\do
    {\@citea\def\@citea{--\penalty\@m}\@ifundefined
       {b@\@citeb}{{\bf ?}\@warning
       {Citation `\@citeb' on page \thepage \space undefined}}%
\hbox{\csname b@\@citeb\endcsname}}}{#1}}
\catcode`@=12

\begin{document}
  \thispagestyle{empty}
  \pagestyle{empty}
  \renewcommand{\thefootnote}{\fnsymbol{footnote}}
\newpage\normalsize
    \pagestyle{plain}
    \setlength{\baselineskip}{4ex}\par
    \setcounter{footnote}{0}
    \renewcommand{\thefootnote}{\arabic{footnote}}
\newcommand{\preprint}[1]{%
  \begin{flushright}
    \setlength{\baselineskip}{3ex} #1
  \end{flushright}}
\renewcommand{\title}[1]{%
  \begin{center}
    \LARGE #1
  \end{center}\par}
\renewcommand{\author}[1]{%
  \vspace{2ex}
  {\Large
   \begin{center}
     \setlength{\baselineskip}{3ex} #1 \par
   \end{center}}}
\renewcommand{\thanks}[1]{\footnote{#1}}
\begin{flushright}
\end{flushright}
\vskip 0.5cm

\begin{center}
{\large \bf Perturbation Foundation of  $q$-Deformed Dynamics}
\end{center}
\vspace{0.4cm}
\begin{center}
Jian-zu Zhang$^{a,b, \S}$
\end{center}
\vspace{0.4cm}
\begin{center}
$^a$ Department of Physics,
University of Kaiserslautern, PO Box 3049, D-67653  Kaiserslautern,
Germany  \\
$^b$ Institute for Theoretical Physics, Box 316,
East China University of Science  and Technology,
Shanghai 200237, P. R. China
\end{center}
\vspace{0.4cm}
\begin{abstract}
In the $q$-deformed theory the perturbation approach can be
expressed in terms of two pairs of  undeformed position and
momentum operators. There are two configuration spaces.
Correspondingly there are two q-perturbation Hamiltonians, one
originates from the perturbation expansion of the potential in one
configuration space, the other one originates from the
perturbation expansion of the kinetic energy in another
configuration space.
 In order to establish a general foundation of
the $q$-perturbation theory,
two perturbation equivalence theorems are proved:
(I) Equivalence theorem {\it I}: Perturbation  expressions of the $q$-deformed
uncertainty relations calculated by two pairs of  undeformed operators
are the same,
and the two $q$-deformed uncertainty relations  undercut
  Heisenberg's minimal one in the same style.
(II) The general equivalence theorem {\it II}:
 for  {\it any} potential (regular or singular)
the expectation values of two q-perturbation Hamiltonians
in the eigenstates
of the undeformed Hamiltonian are equivalent to all orders of the
perturbation expansion.
As an example of singular potentials the  perturbation energy
spectra of the $q$-deformed Coulomb potential are studied.
\end{abstract}
\begin{flushleft}
${^\S}$ E-mail address:  jzzhang@physik.uni-kl.de  \\
\hspace{3.1cm} jzzhangw@online.sh.cn
\end{flushleft}
\clearpage
In searching for new physics at extremely small space scale,
motivated by recent interest of
new field theoretical models and quantum theories of gravity,
there are studies of quantum theories in non-commutative spaces.
The realization of such quantum theories has different approaches.
In one approach the $q$-deformed quantum theory, as a possible
modification of
the ordinary quantum theory at space scale much
smaller than  $10^{-18}$ cm, has attracted attention.
In literature different frameworks of $q$-deformed quantum theories
were established \citer{Schwenk,JZZ02a}. We work in the framework
of the $q$-deformed Heisenberg algebra developed in
Refs.~\cite{Hebecker,Fichtmuller}, which is self-consistent and
 shows interesting physical content.
In this framework characteristics of dynamics and uncertainty relations
of $q$-deformed quantum mechanics are explored \citer{Schwenk,Cerchiai},
\citer{JZZ98,JZZ02a}.

Perturbation $q$-deformed dynamics are involved.
The reason is that there are  two pairs of undeformed variables
 ($\hat x$, $\hat p$) and  ($\tilde x$, $\tilde p),$
and two natural representations of the $q$-deformed operators in
terms of their  undeformed counterparts
\cite{Hebecker,Fichtmuller}. Correspondingly there are two
$q$-perturbation Hamiltonians, one originates from the
perturbation expansion of the potential in  the ($\hat x$, $\hat
p$) system, the other  originates from the perturbation expansion
of the kinetic energy in the ($\tilde x$, $\tilde p$) system
\cite{JZZ98,JZZ00,ZO01,JZZ01}. At the level of operators these two
$q$-perturbation Hamiltonians are different. In the examples of
the harmonic-oscillator potential and the Morse potential,
calculations showed that expectation values of two q-perturbation
Hamiltonians in the eigenstates of the undeformed Hamiltonian are
equivalent \cite{ZO01}. In reference \cite{JZZ01} an equivalence
theorem for regular potentials is demonstrated.

The two pairs of  undeformed variables ($\hat x$, $\hat p$)
and ($\tilde x$, $\tilde p$)  are related by a non-trivial transformation
\cite{Hebecker,Fichtmuller}.
It should be emphasized that this transformation
is not a unitary transformation in a Hilbert space.
Though it maintains the commutation relations $[\hat x$, $\hat p],$
it is not clear whether it leads to the same physical consequences
in general cases.

In order to establish the foundation of the $q$-perturbation theory
in this paper we  demonstrate  two equivalence theorems for general cases.
The equivalence theorem~{\it I}
  states that perturbation expressions of $q$-deformed uncertainty
 relations calculated in the ($\hat x$, $\hat p$) system
and the ($\tilde x$, $\tilde p$) system are the same,
 and the two $q$-deformed uncertainty relations
undercut  Heisenberg's minimal one in the same style.
The equivalence theorem ~{\it II} states that for {\it any}
potential (regular or singular) the expectation values of two
q-perturbation Hamiltonians in the eigenstates of the undeformed
Hamiltonian are equal to all orders  of perturbation
 expressions.
Besides regular potentials demonstrated before \cite{ZO01,JZZ01},
as an example of singular potentials the $q$-deformed
Coulomb potential is studied in detail.

In the following we first review the background.
In terms of the $q$-deformed phase space variables $-$ the
position operator $X$ and the momentum operator $P$, the following
$q$-deformed Heisenberg algebra has been developed \cite{Hebecker,
Fichtmuller}:
\begin{equation}
\label{Eq:q-algebra}
q^{1/2}XP-q^{-1/2}PX=iU, \qquad
UX=q^{-1}XU, \qquad
UP=qPU,
\end{equation}
where $X$ and $P$ are hermitian and $U$ is unitary:
$X^{\dagger}=X$, $P^{\dagger}=P$, $U^{\dagger}=U^{-1}$.
Compared to the Heisenberg algebra the operator $U$ is a new member,
called scaling operator.
Necessity of introducing the operator $U$ is as follows.

Simultaneous hermitian of  $X$ and  $P$ is a delicate point in
the $q$-deformed dynamics.
The definition of the algebra (\ref{Eq:q-algebra}) is based on the
definition of the hermitian momentum operator $P$. However, if $X$
is assumed to be a hermitian operator in a Hilbert space, the
$q$-deformed derivative \cite{Wess}
\begin{equation*}
\partial_X X=1+qX\partial_X,
\nonumber
\end{equation*}
which codes the non-commutativity of space, shows that
 the usual quantization rule $P\to -i\partial_X$ does not yield a
hermitian momentum operator. A hermitian momentum operator $P$ is
related to $\partial_X$ and $X$ in a nonlinear way by introducing
a scaling operator $U$ \cite{Fichtmuller}
\begin{equation*}
U^{-1}\equiv q^{1/2}[1+(q-1)X\partial_X], \qquad
\bar\partial_X\equiv -q^{-1/2}U\partial_X, \qquad
P\equiv -\frac{i}{2}(\partial_X-\bar\partial_X),
\end{equation*}
where $\bar\partial_X$ is the conjugation of $\partial_X$. The
operator $U$ is introduced in the definition of the hermitian
momentum, thus it closely relates to properties of dynamics and
plays an essential role in the $q$-deformed quantum mechanics.
Non-trivial properties of $U$ imply that the algebra
(\ref{Eq:q-algebra}) has a richer structure than Heisenberg's
commutation relation. In the algebra (\ref{Eq:q-algebra}) the
parameter $q$ is a fixed real number. It is important to
distinguish  different realizations of the $q$-algebra by different
ranges of $q$ values \citer{Zachos,Solomon}. Following
Refs.~\cite{Hebecker,Fichtmuller}  we only consider the case $q>1$
 in this paper. The reason is that such choice of  the parameter $q$
leads to a consistent dynamics. In the limit $q\to 1^+$ the scaling
operator $U$ reduces to the unit operator, thus the  algebra
(\ref{Eq:q-algebra}) reduces to Heisenberg's commutation relation.
Such defined hermitian momentum $P$ leads to $q$-deformation
effects, which exhibit in dynamical equations. The  momentum
$P$  non-linearly depends  on $X$ and $\partial_X$. Thus the
$q$-deformed Schr\"odinger equation is difficult to treat.

The $q$-deformed phase space variables $X$, $P$ and
the scaling operator $U$ can be realized in terms of two pairs of
 undeformed variables \cite{Fichtmuller}.

(I) The variables $\hat x$, $\hat p$ of the  ordinary quantum mechanics,
where $\hat x$, $\hat p$ satisfy:
$[ \hat x, \hat p ]=i$, $\hat x=\hat x^{\dagger}$,
 $\hat p=\hat p^ {\dagger}.$
The $q$-deformed operators $X$, $P$ and $U$ are related to
$\hat x$, $\hat p$ as follows:
\begin{equation}
\label{Eq:P-p}
X= \frac{[\hat z+\half]}{\hat z+\half}\hat x,  \qquad
P=\hat p, \qquad
U= q^{\hat z}, \qquad    \hat z=-\frac{i}{2}(\hat x\hat p+\hat p\hat x)
\end{equation}
where $[A]$ is the $q$-deformation of $A$, defined by
$[A]\equiv (q^A-q^{-A})/(q-q^{-1})$.
 It is easy to check that $X$, $P$ and $U$ satisfy the algebra
(\ref{Eq:q-algebra}).

(II) The variables $\tilde x$ and $\tilde p$ of an undeformed algebra,
which are obtained by a  transformation of $\hat x$ and $\hat p$:
\begin{equation}
\label{Eq:tilde}
\tilde x=\hat x F^{-1}(\hat z), \qquad \tilde p= F(\hat z)\hat p, \qquad
F^{-1}(\hat z)= \frac{[\hat z-\half]}{\hat z-\half}.
\end{equation}
Such defined variables $\tilde x$ and $\tilde p$  also satisfy undeformed
algebra: $[ \tilde x, \tilde p ]=i$, and
$\tilde x=\tilde x^{\dagger}$,\quad$\tilde p=\tilde p^{ \dagger}$.
Thus $\tilde p=-i\partial_{\tilde x},$ \;
where $\partial_{\tilde x}{\tilde x}\equiv 1.$
The $q$-deformed operators $X$, $P$ and $U$ are related to
$\tilde  x$ and $\tilde p$ as follows:
\begin{equation}
\label{Eq:X-x}
X=\tilde x, \qquad P=F^{-1}(\tilde z) \tilde p, \qquad
U= q^{\tilde z}, \qquad  \tilde z=-\frac{i}{2}(\tilde x\tilde p +
\tilde p\tilde x),
\end{equation}
where $F^{-1}(\tilde z)$ is defined by Eq.~(\ref{Eq:tilde}) for variables
($\tilde  x$, $\tilde p$).
From Eqs.~(\ref{Eq:tilde}) and (\ref{Eq:X-x}) it follows that such defined
$X$,
$P$ and $U$ also satisfy algebra (\ref{Eq:q-algebra}), and
 Eq.~(\ref{Eq:X-x})  is equivalent to Eq.~(\ref{Eq:P-p}).

The $q$-deformed phase space ($X$, $P$) governed by the $q$-algebra
(\ref{Eq:q-algebra}) is a $q$-deformation of the phase space
($\hat x$, $\hat p$) of the ordinary quantum mechanics,
 thus all machinery of the ordinary quantum mechanics can be applied to
the $q$-deformed quantum mechanics.
It means that dynamical equations of a quantum system are the same
for the undeformed phase space variables ($\hat x$, $\hat p$),
($\tilde x$, $\tilde p$) and for
the $q$-deformed  phase space variables ($X$, $P$), that is, the
$q$-deformed Hamiltonian with the potential $V(X)$ is
$H(X,P)=P^{2}/(2\mu)+V(X).$

Now we consider perturbation treatment of this $q$-deformed theory.
In view of every success of the ordinary quantum
mechanics the  effects of the $q$-deformation must be extremely small,
thus the  perturbation investigation of the $q$-deformed dynamics is
 meaningful,
and the parameter $q$ must be extremely close to one.
So we can  let $q=e^{f}=1+f$, with $0<f\ll 1.$
It is enough accurate to the order $f^2$ in the perturbation treatment.

In the  ($\hat x$, $\hat p$) system and the  ($\tilde x$, $\tilde p$) system
from Eq.~(\ref{Eq:P-p}) and  Eq.~(\ref{Eq:X-x}),
to the order $f^2,$ it follows that the perturbation expansions of  $X$
and $P$ are
\begin{equation}
\label{Eq:X-perturb} X=\hat x  + f^2 g(\hat x, \hat p), \qquad
g(\hat x,\hat p)=-\frac{1}{6}(1+ \hat x  \hat p  \hat x \hat p
)\hat x.
\end{equation}
\begin{equation}
\label{Eq:P-perturbation} P=\tilde p  + f^2 h(\tilde x ,\tilde
p),\qquad h(\tilde  x,\tilde p) =-\frac{1}{6}(1+ \tilde  p\tilde
x\tilde  p\tilde x )\tilde p.
\end{equation}

The operator  $F^{-1}(\hat z)$ defined by Eq.~(\ref{Eq:tilde})
 is not  unitary,
$F^{-1}(\hat z)\ne F^{\dagger}(\hat z)$,
which is a variable transformation between  two configuration spaces;
should be distinguished from a unitary transformation in a
Hilbert space.
It is not clear whether two perturbation formulations in
the $(\hat x,\hat p)$ system and the  $(\tilde x,\tilde p)$ system are
equivalent.
The situation is clarified by the following two equivalence theorems.

First we consider the perturbation treatment of the $q$-deformed
uncertainty relation.

{\bf Perturbation Equivalence Theorem~{\it I}}: The perturbation expressions of
the $q$-deformed uncertainty relation calculated in the $(\hat x,\hat p)$
system and
the $(\tilde x,\tilde p)$ system are the same.

From the algebra (\ref{Eq:q-algebra}) we obtain
\begin{equation*}
XP - PX = iG, \quad G = (U + U^{\dagger})/(q^{1/2} + q^{-1/2}).
\end{equation*}
To the order $f^2$ of the perturbation expansions  in the $(\hat
x,\hat p)$ system and the $(\tilde x,\tilde p)$ system the
operator $G$ has the same representation:
$G=1-\frac{1}{2}f^2\xi\kappa\xi\kappa$, where and in the follows
$(\xi,\kappa)$ represents $(\hat x,\hat p)$ or $(\tilde  x,\tilde
p).$ The corresponding $q$-deformed uncertainty relation reads
\begin{equation}
\label{Eq:ur}
\Delta X\cdot\Delta P \ge \frac{1}{2}|<G>|\ge \frac{1}{2}-
\frac{1}{4} f^2|<\xi\kappa\xi\kappa>|.
\end{equation}

{\bf Undercutting Phenomenon.} The equivalence theorem {\it I} shows that
the $q$-deformed uncertainty relation  essentially deviates from the
 Heisenberg one:
for the case $\Delta X\cdot\Delta P = \frac{1}{2}-
\frac{1}{4} f^2|<\xi\kappa\xi\kappa>|$
the Heisenberg minimal uncertainty relation
$\Delta X\cdot\Delta P = \frac{1}{2}$ is undercut in the same style in the
two perturbation formulations.

Now we consider the perturbation treatment of singular potentials.
As an example, we study the Coulomb potential in detail.

In the $(\hat x,\hat p)$ system the definition of the $q$-deformed
Coulomb potential is involved. Here we give its perturbation
definition.
Because of $f\ll 1$ we have $f^2||g(\hat x,\hat p)||<|\hat x|$
where $||A||$ is the norm of the operator $A.$
In the perturbation
expansion, to the order $f^2,$ the $q$-deformed Coulomb potential is
defined as
\begin{eqnarray}
V(X)=\left\{\begin{array}{ll}
-\kappa/\left[\hat x+f^2 g(\hat x,\hat p)\right] &\textrm{if $\hat x>0$}\\
-\kappa/\left[-\hat x+f^2 g(-\hat x,-\hat p)\right] &\textrm{if
$\hat x<0$}
\end{array} \right.
\label{Eq:Coulomb1}
\end{eqnarray}
where $\kappa>0.$ In the limit $q\to 1^{+}$ the above $q$-deformed
Coulomb potential reduces to the undeformed one $V(\hat x)=-\kappa
|\hat x|^{-1}.$
For singular potentials
 we use the following operator equation to treat the perturbation
expansion:
\begin{equation*}
\frac{1}{A+B}=\frac{1}{A}-\frac{1}{A}B\frac{1}{A}+
\frac{1}{A}B\frac{1}{A}B\frac{1}{A}
-\frac{1}{A}B\frac{1}{A}B\frac{1}{A}B\frac{1}{A}+\cdots,
\nonumber
\end{equation*}
where the norms of operators $A$ and $B$ satisfy $\|B\|<\|A\|.$
Using  Eq.~(\ref{Eq:P-p}) and carefully considering the ordering between
the non-commutative quantities
$ \hat x$ and $g(\hat x,\hat p)$  in the perturbation
 expansion, to the order $f^2$,
we express the $q$-deformed Hamiltonian of the
Coulomb system by the undeformed variables ($\hat x$, $\hat p$) as
 $H(X,P)=H_{un}(\hat x,\hat p) +\hat H^{(q)}_{I,C}(\hat x,\hat p),$
 where the perturbation Hamiltonian
\begin{eqnarray}
\hat H^{(q)}_{I,C}(\hat x,\hat p)=\left\{\begin{array}{ll} \hat
H^{(q)}_{I+}(\hat x,\hat p)
&\textrm{if $\hat x>0$}\\
\hat H^{(q)}_{I-}(\hat x,\hat p)
 &\textrm{if $\hat x<0$}
\end{array} \right.
\label{Eq:Coulomb2}
\end{eqnarray}
and
\begin{equation}
\label{Eq:Hi-hat} \hat H^{(q)}_{I+}(\hat x,\hat p)
=-\frac{1}{6}\kappa f^2(\frac{1}{\hat x}-i\hat p+\hat x\hat p^2),
\;(\hat x>0); \qquad  \hat H^{(q)}_{I-}(\hat x,\hat p)= \hat
H^{(q)}_{I+}(-\hat x,-\hat p), \;(\hat x<0).
\end{equation}

In the  ($\tilde x$, $\tilde p$) system the $q$-deformed
potentials have the same representations as the undeformed ones,
$V(X)=V(\tilde x)=-\kappa/|\tilde x|.$ But the momentum operator
$P$ is a nonlinear function of ($\tilde x$, $\tilde p$). Using
Eq.~(\ref{Eq:X-x}) and carefully considering the ordering between
the non-commutative quantities $ \tilde p$ and $h(\tilde x,\tilde
p)$ in the perturbation expansion, to the order $f^2,$ it follows
that the $q$-deformed Hamiltonian $H(X,P)=H_{un}(\tilde x,\tilde
p)+ \tilde H^{(q)}_{I,C}(\tilde x,\tilde p),$ where the
perturbation Hamiltonian is
\begin{equation}
\label{Eq:Hi-tilde} \tilde H^{(q)}_{I,C}(\tilde x,\tilde p) =
 -\frac{f^2}{12\mu} \bigl[
2\tilde x^2 \tilde p^4- 8i\tilde x  \tilde p^3- 3\tilde p^2 \bigr].
\end{equation}
In the above the undeformed Hamiltonian is
$H_{un}(\xi,\rho)=\rho^2/(2\mu)-\kappa/|\xi|.$

The two perturbation Hamiltonians $\hat H^{(q)}_{I,C} (\hat x,\hat
p)$ and $\tilde H
^{(q)}_{I,C} (\tilde x,\tilde p)$ originate,
separately, from the perturbation expansions  of the potential and
the kinetic energy. At the level of operator they are different.
 Now we show that their contributions to the perturbation
 shifts of the energy spectrum of the undeformed  Hamiltonian in the
 $(\hat x,\hat p)$ system and the $(\tilde  x,\tilde p)$  system  are the
same.

As is well known that for the undeformed one-dimensional Coulomb
system \cite{Loudon} all the excited bound states are  twofold
degenerate, having an even and an odd wave function for each
eigenvalue, except for the ground state which is an even state
localized at the point $\hat x=0$ and having infinite binding
energy. The even state $\psi_{n+}$ and the odd state $\psi_{n-}$ are:
\begin{eqnarray}
\label{Eq:state} \psi_{n\pm}(\hat x)=\left\{\begin{array}{ll}
\psi_{n}(\hat x) &\textrm{if $\hat x>0$}\\
\pm \psi_{n}(-\hat x) &\textrm{if $\hat x<0$}
\end{array} \right.
\end{eqnarray}
where
\begin{equation*}
\psi_{n}(\hat x)=\hat x e^{-\hat x /n} F(1-n,2,2\hat x/n),
\nonumber
\end{equation*}
and $F(1-n,2,x)$ is the usual confluent hypergeometric function.

Now we calculate the energy shifts in the $(\hat x,\hat p)$ system
contributed by the Hamiltonian $\hat H^{(q)}_{I,C}(\hat x,\hat p).$
From Eqs.~(\ref{Eq:Coulomb2}),
(\ref{Eq:Hi-hat}) and (\ref{Eq:state}) it follows that for the
even and the odd state the perturbation shifts of
the undeformed spectrum are
\begin{eqnarray}
\label{Eq:Delta-E-hat} \Delta \hat E^{(q)}_n
&=&\int_{-\infty}^\infty d\hat x \psi^{(0)*}_{n\pm}(\hat x) \hat
H^{(q)}_{I,C}(\hat x,\hat p) \psi^{(0)}_{n\pm}(\hat x)
\nonumber\\
&=&\int_{-\infty}^0 d\hat x\Bigl(\pm\psi^{(0)*}_n(-\hat
x)\Bigr)\hat H^{(q)}_{I-}(\hat x,\hat p)
\Bigl(\pm\psi^{(0)}_n(-\hat x)\Bigr)
\nonumber\\
&&+\int_0^{\infty} d\hat x \psi^{(0)*}_n(\hat x)\hat
H^{(q)}_{I+}(\hat x,\hat p) \psi^{(0)}_n(\hat x)
\nonumber\\
 &=&2\int_0^{\infty} d\hat x \psi^{(0)*}_n(\hat
x)\hat H^{(q)}_{I+}(\hat x,\hat p)\psi^{(0)}_n(\hat x)
\nonumber\\
&=& -\frac{\kappa f^2}{3}\int_0^{\infty} d\hat x
\psi^{(0)*}_n(\hat x)\Bigl\{\frac{1}{\hat x}-i\hat p+\hat x\hat
p^2\Bigr\}\psi^{(0)}_n(\hat x)
\end{eqnarray}

Similarly, in the $(\tilde x,\tilde p)$ system the energy shifts
contributed by the Hamiltonian $\tilde H^{(q)}_{I,C}(\tilde x,\tilde p)$
in Eq.~(\ref{Eq:Hi-tilde}) are
\begin{eqnarray}
\label{Eq:Delta-E-tilde} \Delta \tilde E^{(q)}_n
&=&\int_{-\infty}^\infty d\tilde x \psi^{(0)*}_{n\pm}(\tilde x)
\tilde H^{(q)}_{I,C}(\tilde x,\tilde p)\psi^{(0)}_{n\pm}(\tilde x)
\nonumber\\
&=&2\int_0^\infty d\tilde x \psi^{(0)*}_n(\tilde x) \tilde
H^{(q)}_{I,C}(\tilde x,\tilde p)\psi^{(0)}_n(\tilde x)
\nonumber\\
&=&-\frac{f^2}{6\mu}\int_0^\infty d\tilde x \psi^{(0)*}_n(\tilde
x) \Bigl\{2\tilde x^2 \tilde p^4- 8i\tilde x \tilde p^3
-3\tilde p^2\Bigr\} \psi^{(0)}_n(\tilde x).
\end{eqnarray}
In the  undeformed stationary states $|\psi^{(0)}\rangle$ the time
derivative of the expectation of the operator $\xi^m\rho^n$ is
\begin{equation*}
\nonumber
 i\frac{d}{dt} \langle \psi^{(0)}|\xi^m\rho^n|\psi^{(0)}\rangle
 = \langle \psi^{(0)}|\left[ \xi^m\rho^n,
\frac{1}{2\mu} \rho^2+V(\xi )\right]|\psi^{(0)}\rangle=0.
\end{equation*}
For the case $m+n=even$ the above equation reduces to
\begin{equation}
\int_0^\infty d\xi \psi^{(0)*}_n(\xi)\left[ \xi^m\rho^n,
\frac{1}{2\mu} \rho^2+V(\xi )\right]\psi^{(0)}_n(\xi)=0.
\label{Eq:Virial2}
\end{equation}
From Eq.~(\ref{Eq:Virial2})  for the cases
of $m=n=3$ and $m=n=2$ it follows that for the Coulomb potential we have
\begin{eqnarray}
&&\int_0^\infty d\xi \psi^{(0)*}_n(\xi)\xi^2 \rho^4\psi^{(0)}_n(\xi)
\nonumber\\
&&=\int_0^\infty d\xi \psi^{(0)*}_n(\xi)\left[i\xi \rho^3
+\kappa\mu \left(\xi \rho^2+2i\rho-\frac{2}{\xi} \right)
\right]\psi^{(0)}_n(\xi). \nonumber
 \nonumber
\end{eqnarray}
\begin{equation*}
\int_0^\infty d\xi \psi^{(0)*}_n(\xi)\xi \rho^3\psi^{(0)}_n(\xi)
=\int_0^\infty d\xi \psi^{(0)*}_n(\xi)\left[\frac{i}{2}\rho
+\kappa\mu \left(\rho+\frac{i}{\xi} \right)
\right]\psi^{(0)}_n(\xi). \nonumber
\end{equation*}
Using the above two equations we prove that Eqs.~(\ref{Eq:Delta-E-hat}) and
(\ref{Eq:Delta-E-tilde}) are equivalent.

In general cases  such equivalence is summarized as

{\bf Perturbation Equivalence Theorem~{\it II}}:
For {\it any} potential (regular or singular) the expectation value
$\Delta \hat E^{(q)}_n$
 of the  Hamiltonian $\hat H^{(q)}_I (\hat x,\hat p)$
and the expectation value $\Delta \tilde E^{(q)}_n$ of the Hamiltonian
$\tilde H^{(q)}_I (\tilde x,\tilde p)$
 in the same eigenstate of the undeformed  Hamiltonian are equal
to the all orders of perturbation expansions. Where $\hat
H^{(q)}_I (\hat x,\hat p)$ originates from the perturbation
expansion of the potential in the $(\hat x,\hat p)$ system;
 $\tilde H^{(q)}_I (\tilde x,\tilde p)$
originates from the perturbation expansion of the kinetic energy
in the $(\tilde x,\tilde p)$ system.

Suppose that the Schr\"odinger equation for the undeformed system
$H_{un}$ is solved, $H_{\rm un}|\psi_n^{(0)}\rangle= E^{\rm
(un)}_n|\psi_n^{(0)}\rangle.$ It is obvious that the structure of
the undeformed wave function $\psi_n^{(0)}(\hat x_0)=\langle \hat
x_0|\psi_n^{(0)}\rangle$ in the configuration space $\hat x_0$ and
the structure of the undeformed wave function $\psi_n^{(0)}(\tilde
x_0)=\langle \tilde x_0|\psi_n^{(0)}\rangle$ in the configuration
space $\tilde x_0$ are the same. Because of the hermitian of
$H_{un}(\xi,\rho)$  it is natural to assume that its eigne wave
functions satisfy the completeness relations $\int |\xi\rangle
d\xi\langle \xi|=I$ in either configuration space
 $\xi=\hat x_0$ or $\xi=\tilde x_0.$

Now the demonstration of the equivalence theorem {\it II} is
simple. In the $(\hat x,\hat p)$ system $H(X,P)=H_{un}(\hat x,\hat
p)+\hat H^{(q)}_I(\hat x,\hat p)$ where the $q$-perturbation
Hamiltonian $\hat H^{(q)}_I (\hat x,\hat p)\equiv V(X(\hat x, \hat
p)) -V(\hat x)$
for any potential (regular or singular).
Taking the expectation value of
$H(X,P)$ in the undeformed state $|\psi_n^{(0)}\rangle,$
we have
$$\langle \psi_n^{(0)}|H(X,P)|\psi_n^{(0)}\rangle =E^{\rm
(un)}_n+\langle \psi_n^{(0)}|\hat H^{(q)}_I(\hat x,\hat
p)|\psi_n^{(0)}\rangle.$$
For the second term in the right hand
side of this equation projecting $|\psi_n^{(0)}\rangle$  to the
base $|\hat x_0\rangle$ and using the completeness relation
$\int |\hat x_0\rangle d\hat x_0 \langle \hat x_0|=I,$
it leads to
$$\int d\hat x_0 \langle \psi_n^{(0)}|\hat x_0\rangle \langle \hat x_0|
 \hat H^{(q)}_I(\hat x,\hat p)|\psi_n^{(0)}\rangle= \int d\hat x_0
\psi_n^{(0)\ast}(\hat x_0 ) \hat H^{(q)}_I(\hat
x_0,-i\partial_{\hat x_0} )\psi_n^{(0)}(\hat x_0 ).$$
Thus we obtain
\begin{eqnarray}
E_n &=& \langle \psi_n^{(0)}|H(X,P)|\psi_n^{(0)}\rangle =E^{\rm
(un)}_n+\Delta \hat E^{(q)}_n, \label{Eq:En1}
\\
\Delta \hat E^{(q)}_n &=& \int d\hat x_0 \psi_n^{(0)\ast}(\hat x_0
) \hat H^{(q)}_I(\hat x_0,-i\partial_{\hat x_0} )\psi_n^{(0)}(\hat
x_0 ).
\label{Eq:En2}
\end{eqnarray}
In the $(\tilde  x,\tilde p)$ system $H(X,P)=H_{un}(\tilde
x,\tilde p)+\tilde H^{(q)}_I(\tilde x,\tilde p)$
 where the $q$-perturbation Hamiltonian
$\tilde H^{(q)}_I (\tilde x,\tilde p)\equiv
\frac{1}{2\mu}P^2((\tilde x,\tilde p))-\frac{1}{2\mu}\tilde p^2.$
By the similar procedure we obtain
\begin{eqnarray}
E_n &=& \langle \psi_n^{(0)}|H(X,P)|\psi_n^{(0)}\rangle =E^{\rm
(un)}_n+\Delta \tilde E^{(q)}_n, \label{Eq:En3}
\\
\Delta \tilde E^{(q)}_n &=& \int d\tilde x_0
\psi_n^{(0)\ast}(\tilde x_0 ) \tilde H^{(q)}_I(\tilde
x_0,-i\partial_{\tilde x_0} )\psi_n^{(0)}(\tilde x_0 ).
\label{Eq:En4}
\end{eqnarray}
From Eqs.~(\ref{Eq:En1}) to (\ref{Eq:En4}) we conclude that to the
all orders of perturbation expansions
\begin{equation}
\Delta \hat E^{(q)}_n = \Delta \tilde E^{(q)}_n.
\label{Eq:En}
\end{equation}

In the above the perturbation Hamiltonian $\tilde H^{(q)}_I
(\tilde x,\tilde p)$ itself is potential independent, for any
potential it keeps the same representation,
but the undeformed wave functions $\psi_n^{(0)}(\tilde x_0)$ are
potential dependent, thus the $q$-perturbation shifts
$\Delta\tilde E^{(q)}_n$ of the undeformed energy spectrum
 in the $(\tilde x,\tilde p)$ system are potential dependent.

In the $q$-deformed quantum theory, unlike the  ordinary quantum theory,
there is a  non-trivial transformation among two pairs of the
undeformed variables $(\hat x,\hat p)$ and $(\tilde x,\tilde p).$
It is not a unitary transformation in a Hilbert space.
Such variable transformation leads to two formulations in two
configuration spaces.
The $q$-perturbation quantum theory is much complex than
the ordinary one.
The equivalence theorems {\it I} and {\it II} clarify  the
 foundation for perturbation calculations in the $q$-deformed dynamics.
Based on the equivalence theorems the perturbation effects can be
calculated in the $(\hat x,\hat p)$ system or the $(\tilde x,\tilde p)$
system.
In the $(\tilde x,\tilde p)$ system for any potential
 the perturbation Hamiltonian  $\tilde H^{(q)}_I(\tilde x,\tilde p)$
   keeps the same form, thus
it provides a unified formulation for calculating the
$q$-perturbation shifts of the energy spectrum.

If the $q$-deformed quantum theory is a relevant theory for extremely short
space scale,
its corrections to the ordinary quantum theory must be extremely small
 in the energy range of nowadays experiments.
 Perturbation studies of the $q$-deformed dynamics shows clear indication
of $q$-deformed modifications to the ordinary quantum theory. The
investigation in the $q$-squeezed state \cite{OZ00}  may provide
some evidence about such $q$-deformed effects to nowadays
experiments. Further exploration of the effects of the $q$-deformation
based on the $q$-deformed equivalence theorems is in
progress.

\vspace{0.4cm}
  This work has been supported by the Deutsche Forschungsgemeinschaft
(Germany).  The author would like to thank  W. R\"uhl
 and  P. Osland for stimulating discussions.
 His work has also been supported by the National Natural Science
Foundation of China under the grant number 10074014 and by the Shanghai
Education Development Foundation.

\clearpage

\end{document}